\documentclass[reprint,amsmath,amssymb,aps,showpacs]{revtex4-1}
\usepackage{graphicx}
\usepackage{amsmath}
\usepackage{subfigure}
\usepackage{mathtools} 
\usepackage{color}
\usepackage{overpic}
\usepackage[export]{adjustbox}
\usepackage{mwe, tikz}
\usepackage{rotating}
\usepackage{latexsym}
\usepackage{csquotes}
\usepackage{tikz} 
\usetikzlibrary{calc,arrows,decorations.markings} 
\usepackage{pict2e}
\usepackage{pgfplots}
\usepackage{float}
\usepackage{hyperref}
\usepackage{xr}
\usepackage{rotating}

\begin{document}
\title{Self-diffusion in plastic flow of amorphous solids}
\author{Kamran Karimi}
\email{Email: kamran.karimi1@ucalgary.ca}
\affiliation{Department of Physics and Astronomy, University of Calgary, 2500 University Drive NW, Calgary, Alberta, Canada T2N 1N4}
\begin{abstract}
We report on a particle-based numerical study of sheared amorphous solids in the dense slow flow regime. 
In this framework, deformation and flow are accompanied by critical fluctuation patterns associated with the macroscopic plastic response and single particle kinematics.  
The former is commonly attributed to the collective slip patterns that relax internal stresses within the bulk material and give rise to an effective \emph{mechanical} noise governing the latter particle-level process.  
In this work, the \emph{avalanche}-type dynamics between plastic events is shown to have a strong relevance on the self-diffusion of tracer particles in the Fickian regime.
As a consequence, strong size effects emerge in the effective diffusion coefficient that is rationalized in terms of avalanche size distributions and the relevant temporal occurrence. 
\end{abstract}
\maketitle
\section{Introduction}
Plasticity in amorphous solids refers to intense (irrecoverable) shear deformation that the flowing material goes through macroscopically without any crushing or crumbling. 
The plastic flow and deformation are accompanied by intermittent spatio-temporal fluctuation patterns that have been recently described within the context of “yielding transition” \cite{LinPNAS2014}. 
The microscopic basis of the fluctuations relates to the appearance of Eshelby-like events, small-scale rearranging particles that relax internal stress locally but incur long-range elastic-type perturbations in the medium \cite{picard2004elastic}. 
In the absence of thermal fluctuations, local isolated events are initially activated in mechanically-driven systems, but then further instability may be triggered and propagates due to long-range interactions. 
The \emph{non}-local triggering mechanism leads to an \emph{avalanche}-like dynamics that reveal critical scale-free statistics. 
This includes power-law distributions of avalanche size and duration associated with diverging relevant length and/or timescales \cite{FerreroPRL2014, BakPRL1987}. 
In this context, the failure phenomenon may be viewed as a true non-equilibrium transition with universal scaling properties \cite{ParisiPNAS2017}.

Another important microscopic picture pertains to the diffuse nature of particle trajectories within sheared disordered solids, akin to a thermally-assisted process.
Based upon mean-field arguments, the observed diffusivity can be ascribed to the emergence of ``mechanical noise" generated by large relaxation events.
A priori, dynamics of plastic avalanches must have a strong bearing on the diffusion process down at micro scales. 
Within this context, Martens \emph{et al.} used a mesoscopic elasto-plastic model to relate flow-induced heterogeneities and diffusion constant with regard to shearing rate sensitivity and finite size effects \cite{martens2011connecting}.
Using a related numerical model in \cite{LinPNAS2014, lin2014density}, avalanche statistics were shown to be straightly linked to rheological flow properties based upon generic scaling arguments and a formal analogy with the depinning transition.
Similar particle-based simulations revealed system-spanning slip patterns that were argued to govern long-time diffusive behavior \cite{maloney2008evolution, lemaitre2009rate, roy2015rheology, maloney2015avalanches}. 
Recent studies \cite{hwang2016understanding, lechenault2010super} report on the emergence of \emph{anomalous} diffusion (L{\'e}vy flight in particular) in driven amorphous solids that may be qualitatively understood in view of the broadly-distributed mechanical noise released by scale-free relaxation events \cite{lin2016mean}. 
The observed super diffusive dynamics was rationalized within the context of continuous time random walk (CTRW) theory \cite{metzler2000random} taking into consideration a Poissonian temporal process with broad ``jump" size distributions. 

Here in this study, our aim is to build a generic relationship between diffusivity and avalanche dynamics that should not be specific to microscopic constituents and/or interactions.
In this framework, displacement fluctuations are shown to exhibit non-trivial scale-dependent features that could be quantified in terms of the magnitude and \emph{topology} corresponding with individual avalanches.
We provide a mean-filed type prediction for the scaling observations based on the fact that the kinematics of each slip event may be idealized via the notion of ``Eshelby" transformations as elementary mesoscale constituents.
The temporal fluctuations, on the other hand, can be interpreted in terms of a Poisson point process with the occurrence frequency between successive events that shows scaling features. 
These mechanisms lead up to a long-time diffusion-like process that is characterized by a size-dependent diffusion constant.

The organization of the paper is as follows.
In Sec. \ref{sec:ModelandProtocol}, the bi-axial shear setup, packing preparation, driving protocol, and relevant simulation details are discussed.
In Sec. \ref{sec:particleDisplacementStatistics}, we quantify avalanche size fluctuations along with variations in tracer particles displacements.
We use a mean-field level argument to link the two sets of statistics and validate the proposed scaling by numerical data.
The interevent time distributions and associated scaling features will be the subject of Sec. \ref{sec:WaitTimeDistributions}.
In Sec. \ref{sec:diffusiveDynamics}, the results from Sec. \ref{sec:particleDisplacementStatistics} and \ref{sec:WaitTimeDistributions} are integrated to describe the diffusion process that governs the long-term temporal dynamics.  

%
%
%
\section{Model and Protocol} \label{sec:ModelandProtocol}
\begin{figure}
	\begin{center}
		\begin{overpic}[width=0.25\textwidth]{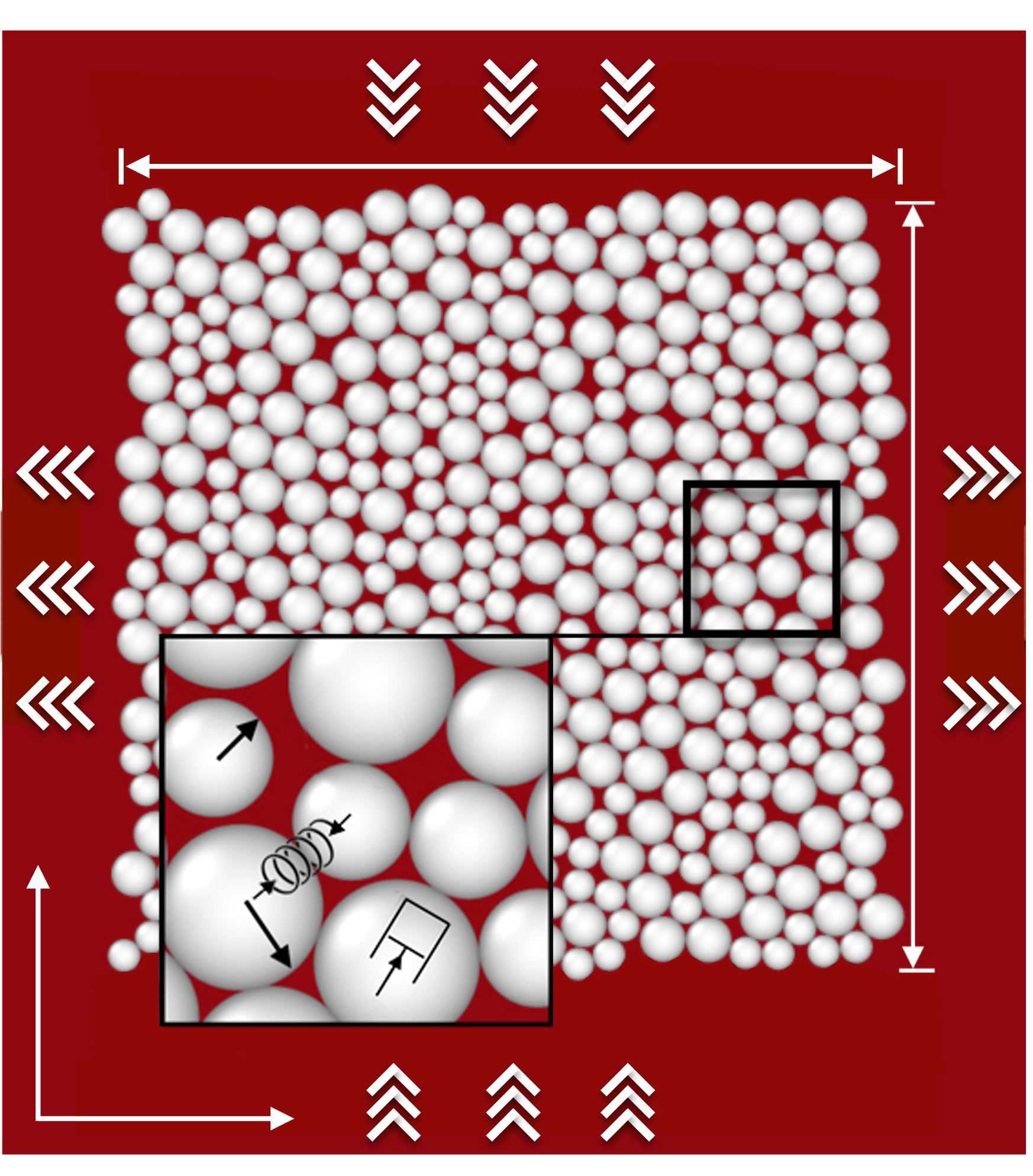}
%
			\put (58,90) {\color{white}$\dot\epsilon_{yy}$} 
             \put (2,65) {\sffamily\setlength{\fboxsep}{0pt}{\strut\bfseries\textcolor{white}{\begin{turn}{90}{$\dot\epsilon_{xx}$}\end{turn}}}} 
			\put (2,28.5) {\color{white}$y$} 
			\put (25.7,3.9) {\color{white}$x$} 
			\put (78,85) {\color{white}\begin{turn}{0}$L$\end{turn}}
			\put (14,35.5) {\begin{turn}{45}$\scriptstyle R_s$\end{turn}} 
			\put (16,22) {\begin{turn}{-44}$\scriptstyle R_b$\end{turn}} 
			\put (29.5,28.5) {\begin{turn}{0}$\scriptstyle k_n$\end{turn}} 
			\put (36,16) {\begin{turn}{0}$\scriptstyle\gamma$\end{turn}} 
		\end{overpic}
	\caption{Bi-axial loading setup. The white discs (with radii $R_s$ and $R_b$) represent the bulk sample with size $L$. 
	The overlapping particles interact via a linear spring $k_{n}$ as sketched in the inset. The dashpot represents the viscous dissipation contribution with drag ratio $\gamma$.  The white arrows indicate the strain-controlled condition with a constant strain rate of $\dot\epsilon_{xx}=-\dot\epsilon_{yy}=\dot\epsilon$.}
	\label{fig:setUP}
	\end{center}
\end{figure}

We used bi-disperse packings of $N$ two dimensional ($d=2$) disks with radii $R_s$ and $R_b$ in a bi-axial loading geometry illustrated in Fig.~\ref{fig:setUP}.
We set ${R_b}/{R_s}=1.4$ and ${N_b}/{N_s}=1$ where $N_{b(s)}$ denotes the number of particles in each species.
The $i\text{-th}$ and $j\text{-th}$ particles with position vectors $\vec{r}_i$, $\vec{r}_j$ may interact with each other when the overlap $\delta=R_i+R_j-|\vec{r}_i-\vec{r}_j|>0$. 
The normal contact forces is $\vec{f}_n=-k_n~\delta~\vec{e}_n$ with the unit normal vector $\vec{e}_n=(\vec{r}_i-\vec{r}_j)/|\vec{r}_i-\vec{r}_j|$.   
Here $k_n$ is the normal spring constant.
A linear drag force $\vec{f}_\text{vis}=-m\tau^{-1}_d~\vec{\dot{r}}$ is applied on each particle with dissipation rate $\tau^{-1}_d$. 
The rate unit (inverse timescale) is set by the vibrational frequency $\omega^2_{n}=k_{n}/m$ where $m$ denotes the particle mass.
Newton's equations of motion were solved in LAMMPS \cite{plimpton1995fast} 
\begin{equation}
m_i\vec{\ddot{u}}_i=\vec{f}_n+\vec{f}_\text{vis}.
\end{equation}
We also set the discretization time $\Delta t=0.05~\omega_n^{-1}$.
An overdamped dynamics was imposed by setting a high value of the damping rate $\tau^{-1}_d$ (in comparison with the vibrational frequency $\omega_n$).

Prior to shearing, samples were prepared by assigning $N$ particles randomly in a bi-periodic $L\times L$ square box with area fraction $\phi=L^{-2}\sum^{N}_{i=1}\pi R_i^2$.
We set $\phi=0.9$, well above the jamming threshold in two dimensional packings.
A strain-controlled condition was then applied by deforming the periodic box along $x$ and $y$ at a constant strain rate $\dot\epsilon_{xx}=-\dot\epsilon_{yy}=\dot\epsilon$. 
The loading protocol was implemented in a discontinuous fashion to ensure the quasi-static condition.
That is, the sample accommodates the incremental strain of $\dot\epsilon\Delta t$ each time step which is followed by a relaxation period with no further deformation (fixed $L$). 
The latter phase terminates once the total kinetic energy $K=\frac{1}{2}\sum^{N}_{i=1}m_i\vec{\dot r}_i.\vec{\dot r}_i < 10^{-10}$ before the next loading period resumes.
We checked that $\dot\epsilon$ was small enough that the stress condition was almost insensitive to the loading rate.

The results of the shear tests may be used to determine the bulk shearing strength together with the structure of deformation during plastic flow. 
The macroscopic stress tensor is defined as 
\begin{equation}
\sigma_{\alpha\beta}=L^{-d}\sum_i\sum_{i<j}(\vec{f}_{ij}\otimes\vec{r}_{ij}) _{\alpha\beta},
\end{equation}
using the Kirkwood-Irvine expression \cite{allen2017computer} where $\vec{f}_{ij}=\vec{f}_n$ and $\vec{r}_{ij}=\vec{r}_i-\vec{r}_j$.
We also compute the non-affine displacement $\vec{u}_i=\vec{u}^\text{~tot}_i-\vec{u}^\text{~aff}_i$ of the $i$-th tracer particle accumulated over time step $\Delta t$ with the total displacement $\vec{u}^\text{~tot}_i=\vec{r}_i(t+\Delta t)-\vec{r}_i(t)$ and affine contribution $\vec{u}^\text{~aff}_i=\dot\epsilon\Delta t(\vec{e}_x\otimes\vec{e}_x-\vec{e}_y\otimes\vec{e}_y)\{\vec{r}_i(t)-\vec{r}_o\}$. 
Here $\vec{r}_o$ and $\vec{e}_{x(y)}$ denote the position vector of the box center and unit vector along $x(y)$, respectively.
%
%
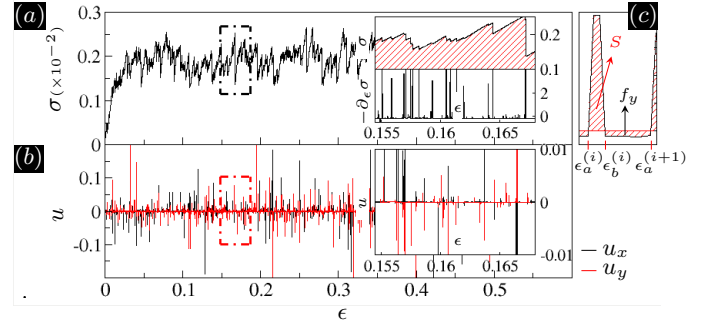
\begin{figure}[t]
	\begin{center} 
		\begin{overpic}[width=8.6cm,]{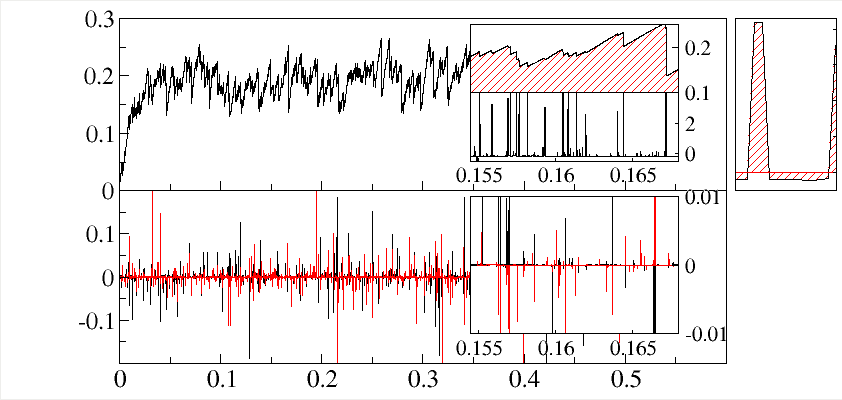}
             \put (50,-2) {$\epsilon$} 
             \put (91,7.9) {$u_x$} 
             \put (91,4.9) {$u_y$} 
             \put (93.5,32) {$\scriptstyle f_y$} 
             \put (86.6,21) {$\scriptstyle \epsilon^{(i)}_a$} 
             \put (91,21) {$\scriptstyle \epsilon^{(i)}_b$} 
             \put (96,21) {$\scriptstyle \epsilon^{(i+1)}_a$} 
             \put (6,13) {\sffamily\setlength{\fboxsep}{0pt}\colorbox{white}{\strut\bfseries\textcolor{black}{\begin{turn}{90}{$u$}\end{turn}}}} 
             \put (5,30) {\sffamily\setlength{\fboxsep}{0pt}\colorbox{white}{\strut\bfseries\textcolor{black}{\begin{turn}{90}{$\sigma$\tiny$(\times10^{-2})$}\end{turn}}}} 
             \put (92,40) {\sffamily\setlength{\fboxsep}{0pt}{\strut\bfseries\textcolor{red}{\begin{turn}{0}{$\scriptstyle S$}\end{turn}}}} 
             \put (53,28) {\sffamily\setlength{\fboxsep}{0pt}{\strut\bfseries\textcolor{black}{\begin{turn}{90}{$\scriptstyle -\partial_\epsilon \sigma$}\end{turn}}}} 
             \put (53,40) {\sffamily\setlength{\fboxsep}{1pt}\colorbox{white}{\strut\bfseries\textcolor{black}{\begin{turn}{90}{$\scriptstyle \sigma$}\end{turn}}}} 
             \put (68,30) {\sffamily\setlength{\fboxsep}{0pt}\colorbox{white}{\strut\bfseries\textcolor{black}{\begin{turn}{0}{$\scriptstyle \epsilon$}\end{turn}}}} 
             \put (68,9.5) {\sffamily\setlength{\fboxsep}{0pt}\colorbox{white}{\strut\bfseries\textcolor{black}{\begin{turn}{0}{$\scriptstyle \epsilon$}\end{turn}}}} 
             \put (53,15) {\sffamily\setlength{\fboxsep}{1pt}\colorbox{white}{\strut\bfseries\textcolor{black}{\begin{turn}{90}{$\scriptstyle u$}\end{turn}}}} 
             \put (0,44) {\sffamily\setlength{\fboxsep}{0pt}\colorbox{black}{\strut\bfseries\textcolor{white}{\small$(a)$}}} 
             \put (0,21.9) {\sffamily\setlength{\fboxsep}{0pt}\colorbox{black}{\strut\bfseries\textcolor{white}{\small$(b)$}}} 
             \put (95,44) {\sffamily\setlength{\fboxsep}{0pt}\colorbox{black}{\strut\bfseries\textcolor{white}{\small$(c)$}}} 
             \begin{tikzpicture}
                 \coordinate (a) at (0,0); 
                 \node[] at (a) {\tiny.};
			     \coordinate (center2) at (8.33,2.08); 
			     \coordinate (center3) at (8.33,1.96); 
                 \coordinate (b) at ($ (center3) + .075*(1,0) $); 
                 \coordinate (c) at ($ (center3) - .075*(1,0) $);
                \draw[red][line width=0.1mm] (center3) -- (center2); 
			     \coordinate (center2) at (7.72,2.08); 
			     \coordinate (center3) at (7.72,1.96); 
                 \coordinate (b) at ($ (center3) + .075*(1,0) $); 
                 \coordinate (c) at ($ (center3) - .075*(1,0) $);
                \draw[red][line width=0.1mm] (center3) -- (center2); 
			     \coordinate (center2) at (7.49,2.08); 
			     \coordinate (center3) at (7.49,1.96); 
                 \coordinate (b) at ($ (center3) + .075*(1,0) $); 
                 \coordinate (c) at ($ (center3) - .075*(1,0) $);
                \draw[red][line width=0.1mm] (center3) -- (center2); 
			     \coordinate (center2) at (7.8,3.2); 
			     \coordinate (center3) at (7.6,2.5); 
                 \coordinate (b) at ($ (center3) + .075*(1,0) $); 
                 \coordinate (c) at ($ (center3) - .075*(1,0) $);
                \draw[->,>=stealth,red][line width=0.1mm] (center3) -- (center2); 
			     \coordinate (center2) at (7.98,2.55); 
			     \coordinate (center3) at (7.98,2.15); 
                 \coordinate (b) at ($ (center3) + .075*(1,0) $); 
                 \coordinate (c) at ($ (center3) - .075*(1,0) $);
                \draw[->,>=stealth,black][line width=0.1mm] (center3) -- (center2); 
                 \coordinate (a) at (0,0); 
                 \node[] at (a) {\tiny.};
                 \coordinate (center2) at (2.6,2.7); 
                 \coordinate (b) at ($ (center2) + .4*(1,0) $);
                 \coordinate (c) at ($ (b) + 0.9*(0,1) $);
                 \coordinate (d) at ($ (c) - 0.4*(1,0) $);
                 \draw[line width=0.3mm,dashdotted] (center2) -- (b); 
                 \draw[line width=0.3mm,dashdotted] (b) -- (c); 
                 \draw[line width=0.3mm,dashdotted] (d) -- (c); 
                 \draw[line width=0.3mm,dashdotted] (d) -- (center2); 
                 \coordinate (a) at (0,0); 
                 \node[] at (a) {\tiny.};
                 \coordinate (center3) at (2.6,.7); 
                 \coordinate (b) at ($ (center3) + 0.4*(1,0) $);
                 \coordinate (c) at ($ (b) + .9*(0,1) $);
                 \coordinate (d) at ($ (c) - 0.4*(1,0) $);
                 \draw[line width=0.3mm,red,dashdotted] (center3) -- (b); 
                 \draw[line width=0.3mm,red,dashdotted] (b) -- (c); 
                 \draw[line width=0.3mm,red,dashdotted] (d) -- (c); 
                 \draw[line width=0.3mm,red,dashdotted] (d) -- (center3); 
 				  \coordinate (a) at (0,0); 
                 \node[] at (a) {\tiny.};
                 \coordinate (center3) at (7.4,.62); 
                 \coordinate (b) at ($ (center3) + 0.2*(1,0) $);
                 \draw[line width=0.1mm] (center3) -- (b); 
                 \coordinate (center3) at (7.4,.35); 
                 \coordinate (b) at ($ (center3) + 0.2*(1,0) $);
                 \draw[line width=0.1mm,red] (center3) -- (b); 
             \end{tikzpicture}
		\end{overpic}
	\end{center}
	\caption{Results of strain-controlled bi-axial tests at $L=80$. Evolution of (\textbf{a}) the bulk shear stress $\sigma$ and (\textbf{b}) the tracer particle displacement $u$ per unit time step along $x$ and $y$ with the imposed strain $\epsilon$. The insets are the close-up views of the main graphs. (\textbf{c}) $-\partial_\epsilon\sigma$ versus $\epsilon$ corresponding to the $i$-th avalanche. The hatched areas over $[\epsilon^{(i)}_a$, $\epsilon^{(i)}_b]$ and $[\epsilon^{(i)}_b$, $\epsilon^{(i+1)}_a]$ denote the avalanche size $S$ and stress threshold $f_y$, respectively. The flat (red) line indicates $\partial_\epsilon\sigma=0$.}
	\label{fig:loadCurve}
\end{figure}
The resulting load curves $\sigma=\frac{1}{2}(\sigma_{xx}-\sigma_{yy})$ against shear strain $\epsilon=\frac{1}{2}(\epsilon_{xx}-\epsilon_{yy})$ along with the non-affine displacements $u$ of a tracer particle along $x$ and $y$ are reported in Fig.~\ref{fig:loadCurve}(a) and (b).
Upon shear loading, the response reveals a well-established steady flow in Fig.~\ref{fig:loadCurve}(a) following the initial yielding regime.  
As evidenced in the upper inset of Fig.~\ref{fig:loadCurve}(a), the stress dynamics is characterized by abrupt falloffs which are preceded by longer periods of stress build-up, an expected feature of amorphous structures.
This bursty dynamics becomes further apparent in $\partial_{\epsilon}\sigma$, the derivative of the stress signal with respect to strain, which is shown in the lower inset of Fig.~\ref{fig:loadCurve}(a). 
Figure~\ref{fig:loadCurve}(b) shows intermittent features that are also present in the tracer particle trajectory and appear to statistically correlate with the extent of stress drops as the deformation proceeds.
The latter indeed relates to the avalanche size $S\doteq-L^d \int_{\epsilon^{(i)}_a}^{\epsilon^{(i)}_b}\partial_{\epsilon}\sigma~d\epsilon$ which has dimensions of energy and corresponds to the $i$-th avalanche incurred at the strain interval $[\epsilon^{(i)}_a,\epsilon^{(i)}_b]$ as sketched in Fig.~\ref{fig:loadCurve}(c). 
Another fluctuating quantity is $f_y\doteq \int_{\epsilon^{(i)}_b}^{\epsilon^{(i+1)}_a}\partial_{\epsilon}\sigma~d\epsilon$ which measures the stored (elastic) energy per volume over the stress accumulation period $[\epsilon^{(i)}_b,\epsilon^{(i+1)}_a]$. 
The stress threshold $f_y$, together with the tracer displacement $u$ and avalanche size $S$, contains non-trivial statistics which will be the subject of the following sections. 

 \section{particle displacement statistics}\label{sec:particleDisplacementStatistics}
The statistical metric we probe is the probability distribution function for plastic avalanches $P(S)$ and particle displacements $P(u)$. 
Due to the loading symmetry along $x$ and $y$, the corresponding displacements are \emph{statistically} equivalent and we, accordingly, drop the subscript for $u$ hereafter.
The statistics are collected independently over multiple sheared samples once a steady-state flow regime was established following an initial transient response, \emph{i.e.} $\epsilon>0.1$.
We used the full set of particles and associated trajectories so as to improve particle-based statistics. 

\begin{figure}[t]
    \begin{center}
        \begin{overpic}[width=8.6cm]{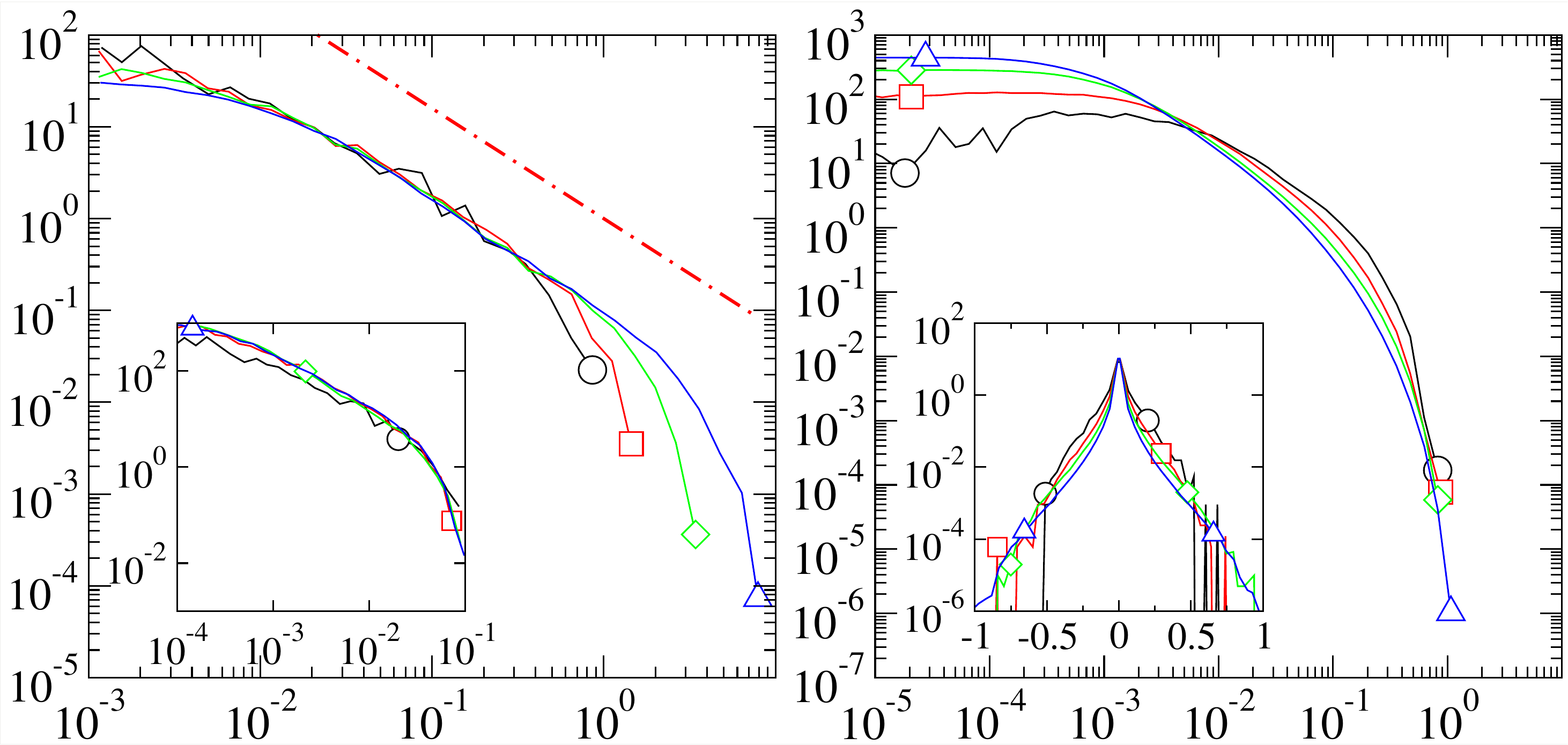}
			 \put (93.5,37.1) {$L$} 
            \put (93.5,34.25) {{\color{black}$\scriptstyle~10$}} 
            \put (93.5,30.8) {{\color{red}$\scriptstyle~20$}} 
            \put (93.5,27.3) {{\color{green}$\scriptstyle~40$}} 
            \put (93.5,23.8) {{\color{blue}$\scriptstyle~80$}} 
            \put (26,-3) {$S$}
            \put (18,10) {$\scriptstyle S/L^{d_f}$}
            \put (76,-3) {$u$} 
            \put (70,10) {$\scriptstyle u$} 
            \put (-4,22) {\sffamily\setlength{\fboxsep}{0pt}\colorbox{white}{\strut\bfseries\textcolor{black}{\begin{turn}{90}{$P(S)$}\end{turn}}}} 
            \put (12,11) {\sffamily\setlength{\fboxsep}{0pt}\colorbox{white}{\strut\bfseries\textcolor{black}{\begin{turn}{90}{$\scriptstyle P(S)L^{\tau.d_f}$}\end{turn}}}} 
            \put (101,22) {\sffamily\setlength{\fboxsep}{0pt}\colorbox{white}{\strut\bfseries\textcolor{black}{\begin{turn}{90}{$P(u)$}\end{turn}}}} 
            \put (63,15) {\sffamily\setlength{\fboxsep}{0pt}\colorbox{white}{\strut\bfseries\textcolor{black}{\begin{turn}{90}{$\scriptstyle P(u)$}\end{turn}}}} 
            \put (44,41.5) {\sffamily\setlength{\fboxsep}{0pt}\colorbox{black}{\strut\bfseries\textcolor{white}{\small$(a)$}}} 
            \put (94,41.5) {\sffamily\setlength{\fboxsep}{0pt}\colorbox{black}{\strut\bfseries\textcolor{white}{\small$(b)$}}} 
           \begin{tikzpicture}
                \coordinate (a) at (0,0); 
                \node[] at (a) {\tiny.};                 
                \coordinate (center1) at (7.8,2.9); 
				 \draw[black] (center1) circle (2pt);
                \coordinate (b) at ($ (center1) - (.075,0) $);
                \coordinate (c) at ($ (b) - (.1,0) $); 
                \coordinate (d) at ($ (center1) + (.075,0) $);
                \coordinate (e) at ($ (d) + (.1,0) $); 
                \draw[black][line width=0.1mm] (b) -- (c); 
                \draw[black][line width=0.1mm] (d) -- (e); 
			     \coordinate (center1) at (7.8,2.61); 
                \coordinate (b) at ($ (center1) - (.075,0) $);
                \coordinate (c) at ($ (b) - (.1,0) $); 
                \coordinate (d) at ($ (center1) + (.075,0) $);
                \coordinate (e) at ($ (d) + (.1,0) $); 
				 \draw[red] ($(b)-(0.0,0.07)$) rectangle ($ (d) + (0.0,0.07) $); 
                \draw[red][line width=0.1mm] (b) -- (c); 
                \draw[red][line width=0.1mm] (d) -- (e); 
			     \coordinate (center1) at (7.8,2.31); 
                \coordinate (b) at ($ (center1) + (0.075,0.0) $); 
                \coordinate (c) at ($ (center1) + (0,0.075) $); 
                \coordinate (d) at ($ (center1) + (-0.075,0) $);
                \coordinate (e) at ($ (center1) + (0.0,-0.075) $);
				 \draw[green] (b) -- (c) -- (d) -- (e) -- (b);
                \coordinate (c) at ($ (b) + (.1,0) $); 
                \coordinate (e) at ($ (d) - (.1,0) $); 
                \draw[green][line width=0.1mm] (b) -- (c); 
                \draw[green][line width=0.1mm] (d) -- (e); 
			     \coordinate (center1) at (7.8,2.01); 
                \coordinate (b) at ($ (center1) - 0.17*(0.5,0.289) $); 
                \coordinate (c) at ($ (center1) + 0.17*(0.5,-0.289) $); 
                \coordinate (d) at ($ (center1) + 0.17*(0,0.366) $);
				 \draw[blue] (b) -- (c) -- (d) -- (b);
                \coordinate (b) at ($ (center1) - (.075,0) $);
                \coordinate (c) at ($ (b) - (.1,0) $); 
                \coordinate (d) at ($ (center1) + (.075,0) $);
                \coordinate (e) at ($ (d) + (.1,0) $); 
                \draw[blue][line width=0.1mm] (b) -- (c); 
                \draw[blue][line width=0.1mm] (d) -- (e); 
				\coordinate (center1) at (3.1,3.2); 
				\coordinate (b) at ($ (center1) - 0.47*(1,0) $);
				\coordinate (c) at ($ (center1) - 0.3*(0,1) $); 
				\draw[red,line width=0.1mm] (center1) -- (b); 
				\draw[red,line width=0.1mm] (center1) -- (c); 
				\draw[red,line width=0.1mm] (b) -- (c); 
				\node at ($(center1)+(0.15,-0.15)$) {{\color{red}$\scriptstyle\tau$}}; 
				\node at ($(center1)+(-0.2,0.1)$) {{\color{red}$\scriptstyle 1$}}; 
			            \end{tikzpicture}
                    \end{overpic}
                            \end{center}
    \caption{Statistics of (\textbf{a}) avalanche sizes $S$ and (\textbf{b}) particle displacements $u$ at multiple system sizes $L=10, 20, 40, 80$. The dashdotted line indicates a power law $S^{-\tau}$ with $\tau\simeq 1.3$. The left inset plots the rescaled distributions $P(S)L^{\tau.d_f}$ versus $S/L^{d_f}$ with $d_f\simeq 1.2$. The right inset is the same as the main graph in (\textbf{b}) but plotted on log-lin scale.}
    \label{fig:pdfSU}
\end{figure}
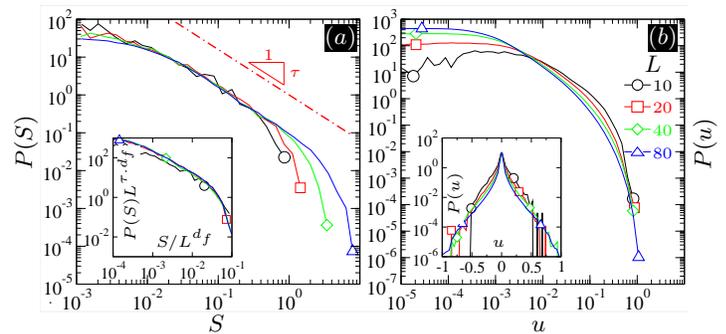

Figure~\ref{fig:pdfSU}(a) points to critical fluctuation patterns that, because of avalanche-type dynamics, occur over a broad range of scales and give rise to power-law statistics over almost three decades.
The exponential-like cut-off in $P(S)$ is a signature of extended system-spanning events that will typically scale with the physical size of the sample.
Statistics of avalanches can be well fit by the empirical scaling form $P(S)\propto S^{-\tau}\text{exp}(-S/L^{d_f})$ as tested in the inset of Fig.~\ref{fig:pdfSU}(a) with $\tau\simeq1.3$ and $d_f\simeq1.2$.
The former exponent is reasonably close to the mean-field prediction of $\tau=\frac{3}{2}$ \cite{lin2014density}.
The latter denotes the fractal dimension quantifying the spatial extension of slip events. 
The preliminary plateau regime corresponds to localized shear modes that would naturally depend on microscopic details rather than linear size $L$.

As for particle displacement distributions in Fig.~\ref{fig:pdfSU}(b), $P(u)$ does not seem to include critical scaling features present in avalanche size statistics.
The data develop a cusp in the center which extends to an exponential decay at intermediate and large $u$ values as illustrated in the inset of Fig.~\ref{fig:pdfSU}(b).
The typical scale (of order unity or particle size) within the exponential tail corresponds with the kinematics of locally rearranging particles 
which is analogous to ``T1" events in foam dynamics \cite{kabla2003local}.  
This upper bound is therefore almost insensitive to the macroscopic size $L$ as reported in \cite{maloney2008evolution}. 
However, size effects are evident in terms of distribution widths with larger samples containing weaker displacement fluctuations.

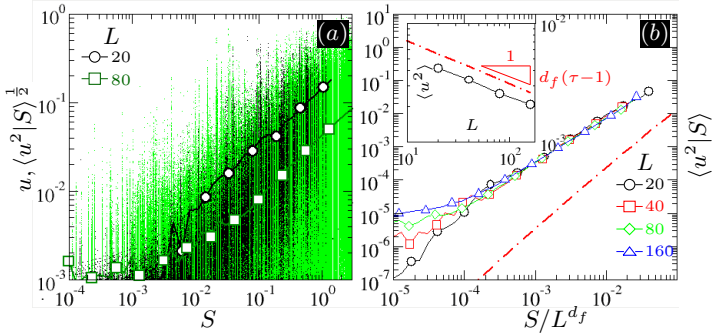
\begin{figure}[t]
    \begin{center}
        \begin{overpic}[width=8.6cm]{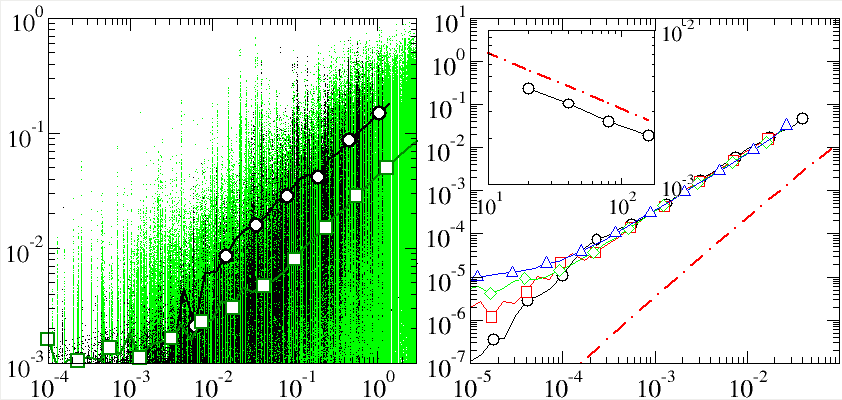}
			 \put (93.5,21.3) {$L$} 
            \put (93.5,18.45) {{\color{black}$\scriptstyle~20$}} 
            \put (93.5,15) {{\color{red}$\scriptstyle~40$}} 
            \put (93.5,11.5) {{\color{green}$\scriptstyle~80$}} 
            \put (93.5,8.0) {{\color{blue}$\scriptstyle~160$}} 
            \put (11,40.85) {$L$} 
            \put (11,38.0) {{\color{black}$\scriptstyle~20$}} 
            \put (11,34.5) {{\color{black!60!green}$\scriptstyle~80$}} 
            \put (67,27.5) {$\scriptstyle L$} 
            \put (26,-3) {$S$}
            \put (76,-3) {$S/L^{d_f}$}  
            \put (-4,18) {\sffamily\setlength{\fboxsep}{0pt}\colorbox{white}{\strut\bfseries\textcolor{black}{\begin{turn}{90}{$u, \langle u^2|S\rangle^{\frac{1}{2}}$}\end{turn}}}} 
            \put (101,20) {\sffamily\setlength{\fboxsep}{0pt}{\strut\bfseries\textcolor{black}{\begin{turn}{90}{$\langle u^2|S\rangle$}\end{turn}}}}  
            \put (59,32) {\sffamily\setlength{\fboxsep}{0pt}{\strut\bfseries\textcolor{black}{\begin{turn}{90}{$\scriptstyle\langle u^2\rangle$}\end{turn}}}}  
            \put (44,41.5) {\sffamily\setlength{\fboxsep}{0pt}\colorbox{black}{\strut\bfseries\textcolor{white}{\small$(a)$}}} 
            \put (94,41.5) {\sffamily\setlength{\fboxsep}{0pt}\colorbox{black}{\strut\bfseries\textcolor{white}{\small$(b)$}}} 
           \begin{tikzpicture}
                \coordinate (a) at (0,0); 
                \node[] at (a) {\tiny.};               %
                \coordinate (center1) at (0.69,3.22); 
				 \draw[black] (center1) circle (2pt);
                \coordinate (b) at ($ (center1) - (.075,0) $);
                \coordinate (c) at ($ (b) - (.1,0) $); 
                \coordinate (d) at ($ (center1) + (.075,0) $);
                \coordinate (e) at ($ (d) + (.1,0) $); 
                \draw[black][line width=0.1mm] (b) -- (c); 
                \draw[black][line width=0.1mm] (d) -- (e); 
			     \coordinate (center1) at (.69,2.93); 
                \coordinate (b) at ($ (center1) - (.075,0) $);
                \coordinate (c) at ($ (b) - (.1,0) $); 
                \coordinate (d) at ($ (center1) + (.075,0) $);
                \coordinate (e) at ($ (d) + (.1,0) $); 
				 \draw[black!60!green] ($(b)-(0.0,0.075)$) rectangle ($ (d) + (0.0,0.075) $); 
                \draw[black!60!green][line width=0.1mm] (b) -- (c); 
                \draw[black!60!green][line width=0.1mm] (d) -- (e); 
                \coordinate (center1) at (7.8,1.52); 
				 \draw[black] (center1) circle (2pt);
                \coordinate (b) at ($ (center1) - (.075,0) $);
                \coordinate (c) at ($ (b) - (.1,0) $); 
                \coordinate (d) at ($ (center1) + (.075,0) $);
                \coordinate (e) at ($ (d) + (.1,0) $); 
                \draw[black][line width=0.1mm] (b) -- (c); 
                \draw[black][line width=0.1mm] (d) -- (e); 
			     \coordinate (center1) at (7.8,1.23); 
                \coordinate (b) at ($ (center1) - (.075,0) $);
                \coordinate (c) at ($ (b) - (.1,0) $); 
                \coordinate (d) at ($ (center1) + (.075,0) $);
                \coordinate (e) at ($ (d) + (.1,0) $); 
				 \draw[red] ($(b)-(0.0,0.07)$) rectangle ($ (d) + (0.0,0.07) $); 
                \draw[red][line width=0.1mm] (b) -- (c); 
                \draw[red][line width=0.1mm] (d) -- (e); 
			     \coordinate (center1) at (7.8,0.93); 
                \coordinate (b) at ($ (center1) + (0.075,0.0) $); 
                \coordinate (c) at ($ (center1) + (0,0.075) $); 
                \coordinate (d) at ($ (center1) + (-0.075,0) $);
                \coordinate (e) at ($ (center1) + (0.0,-0.075) $);
				 \draw[green] (b) -- (c) -- (d) -- (e) -- (b);
                \coordinate (c) at ($ (b) + (.1,0) $); 
                \coordinate (e) at ($ (d) - (.1,0) $); 
                \draw[green][line width=0.1mm] (b) -- (c); 
                \draw[green][line width=0.1mm] (d) -- (e); 
			     \coordinate (center1) at (7.8,0.63); 
                \coordinate (b) at ($ (center1) - 0.17*(0.5,0.289) $); 
                \coordinate (c) at ($ (center1) + 0.17*(0.5,-0.289) $); 
                \coordinate (d) at ($ (center1) + 0.17*(0,0.366) $);
				 \draw[blue] (b) -- (c) -- (d) -- (b);
                \coordinate (b) at ($ (center1) - (.075,0) $);
                \coordinate (c) at ($ (b) - (.1,0) $); 
                \coordinate (d) at ($ (center1) + (.075,0) $);
                \coordinate (e) at ($ (d) + (.1,0) $); 
                \draw[blue][line width=0.1mm] (b) -- (c); 
                \draw[blue][line width=0.1mm] (d) -- (e); 
				\coordinate (center1) at (6.48,3.1); 
				\coordinate (b) at ($ (center1) - 0.65*(1,0) $);
				\coordinate (c) at ($ (center1) - 0.25*(0,1) $); 
				\draw[red,line width=0.1mm] (center1) -- (b); 
				\draw[red,line width=0.1mm] (center1) -- (c); 
				\draw[red,line width=0.1mm] (b) -- (c); 
				\node at ($(center1)+(0.6,-0.18)$) {{\color{red}$\scriptstyle d_f(\tau-1)$}}; 
				\node at ($(center1)+(-0.26,0.14)$) {{\color{red}$\scriptstyle 1$}}; 
				            \end{tikzpicture}
                    \end{overpic}
        \end{center}
    \caption{Statistics of tracer particles motion conditioned on avalanche size $S$ at different sample sizes $L$. (\textbf{a}) Scatter plot of the correlations between displacements $u$ and $S$ at $L=20, 80$. The solid curves show conditional variance $\langle u^2|S\rangle$ plotted against $S$ (\textbf{b}) $\langle u^2|S\rangle$ vs. $S/L^{d_f}$ at $L=10, 20, 40, 80$. The dashdotted line indicates the $\langle u^2|S\rangle\propto-({S}/{L^{d_f}})^2~\text{Log}({S}/{L^{d_f}})$ scaling with $d_f=1.2$. The inset plots the expected variance $\langle u^2\rangle$ against $L$. The dashdotted (red) line is a guide to the power law $\langle u^2\rangle\propto L^{-d_f(\tau-1)}$ with $\tau=1.3$.}
    \label{fig:scatterUS}
\end{figure}

We further examined the root mean squared fluctuations $\langle u^2|S\rangle^{\frac{1}{2}}$ conditioned on avalanche size $S$. 
Apart from the plateau region at small values of $S$, the scatter plot of Fig.~\ref{fig:scatterUS}(a) indicates that incident avalanches with larger magnitudes will, in general, result in a broader \emph{noise} distribution.
We noted similar trends in a model metallic glass where the cross-over behavior was associated with the interplay between small and large events \cite{cao2018nanomechanics}. 
The data collapse in Fig.~\ref{fig:scatterUS}(b) signifies that the width of distributions is uniquely determined by the re-scaled avalanche size $S/L^{d_f}$. 
We provide a mean-filed approximation for the observed scaling behavior associated with displacement fluctuations.
Within this approach, avalanches are effectively treated as quasi-linear objects (with $d_f=1$) in two dimensional space ($d=2$) extended over size $\xi$ within a system of linear size $L$. 
This fractal unit is assumed to be constructed by a set of individual Eshelby elements that incur decaying displacements of the form $u\propto 1/r^{d-1}$ in the elastic medium \cite{karimi2018correlation}.
Superimposing individual contributions, it follows that 
\begin{eqnarray}
u(r|\xi)&=&\int_{x=-\frac{\xi}{2}}^{+\frac{\xi}{2}}\frac{1}{|r-x|^{d-1}}~dx \nonumber \\
&\propto&\text{tanh}^{-1}(\frac{\xi}{r})~~~~~~r>\frac{\xi}{2},
\end{eqnarray}
which shows a slow logarithmic decay in the near-field. 
Note that the $u\propto 1/r$ scaling will be recovered using the far-field approximation at $r\gg\xi$. 

The Displacement distribution conditioned on size $\xi$ will be given as 
\begin{eqnarray}
P(u|\xi)&=&\int_{r=\frac{\xi}{2}}^{L}\delta(u-\text{tanh}^{-1}(\frac{\xi}{r}))~d^dr \nonumber\\
&\propto& \frac{\text{cosh}(u)}{\text{sinh}^3(u)}~~~~~~\text{tanh}^{-1}(\frac{\xi}{L})<u<\infty,
\end{eqnarray}
with the lower cutoff set by the finite system size.
Therefore,  
\begin{eqnarray}
<u^2|\xi>&=&\int u^2P(u|\xi)~du \nonumber\\
&\propto& -(\frac{\xi}{L})^2~\text{Log}(\frac{\xi}{L})+O[(\frac{\xi}{L})^3]~~~~\xi\ll L.
\end{eqnarray}
Upon inserting $(\xi/L)^{d_f}\propto S/L^{d_f}$ and $d_f=1$, the proposed scaling is plotted in Fig.~\ref{fig:scatterUS}(b) 
which appears to slightly overestimate the power-law like growth of fluctuations.

Given the above functional forms, the expected variance $\langle u^2\rangle=\int\langle u^2|S\rangle P(S)~dS$ should now scale as $\langle u^2\rangle\propto 1/L^{d_f(\tau-1)}$.
The derived scaling relation indicates that a reduction in $\tau$ will amplify fluctuations which is meaningful since shallow avalanche distributions imply high occurrence frequencies of large avalanches.
The numerical data agree pretty well with theoretical predictions as in the inset of Fig.~\ref{fig:scatterUS}(b) indicating the strong relevance of stress fluctuation patterns (and associated critical exponents) on single particle statistics.  

\begin{figure}[t]
    \begin{center}
        \begin{overpic}[width=8.6cm]{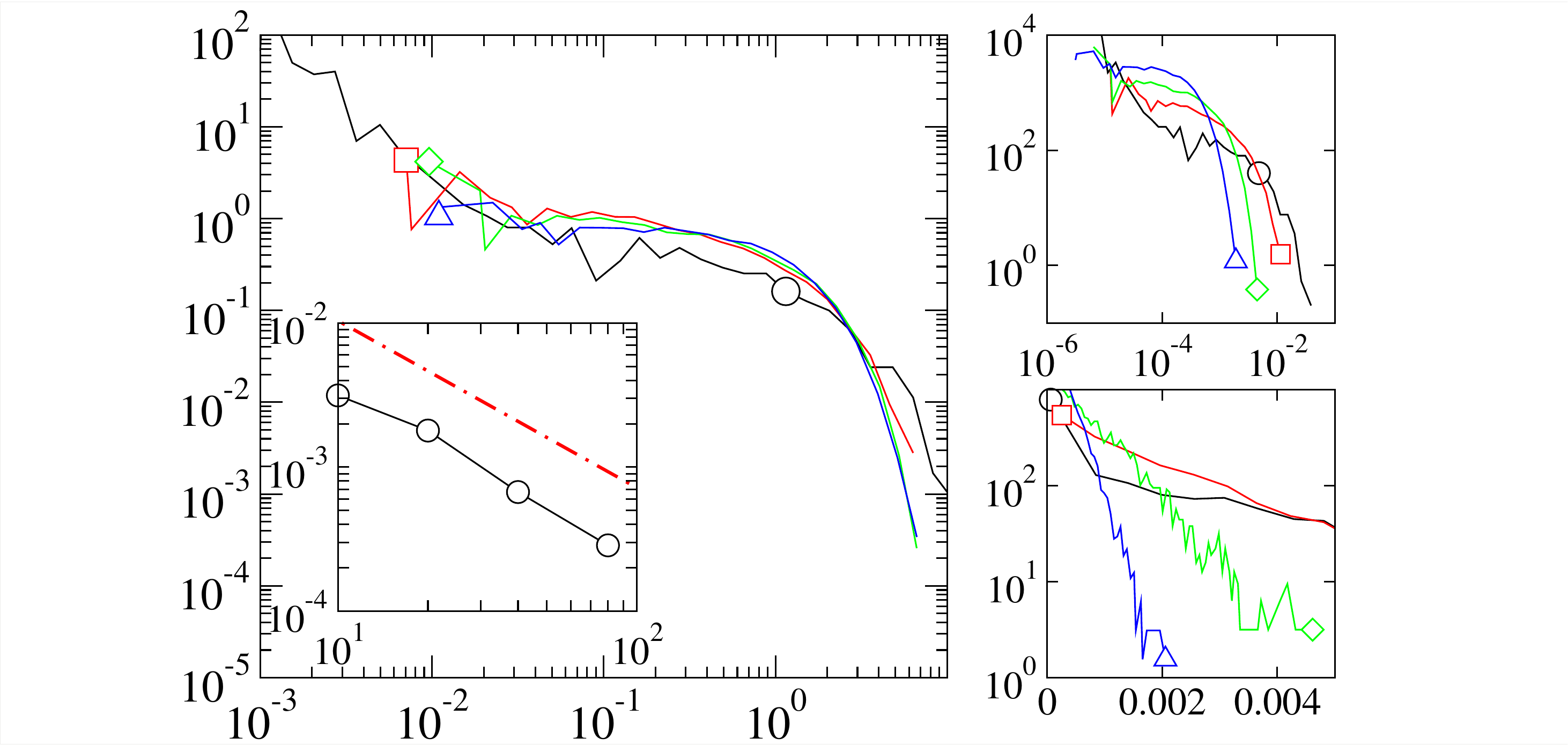}
			 \put (93.5,37.1) {$L$} 
            \put (93.5,34.25) {{\color{black}$\scriptstyle~10$}} 
            \put (93.5,30.8) {{\color{red}$\scriptstyle~20$}} 
            \put (93.5,27.3) {{\color{green}$\scriptstyle~40$}} 
            \put (93.5,23.8) {{\color{blue}$\scriptstyle~80$}} 
            \put (34,-3) {$f_y/\bar{f}_y$}
            \put (30,10) {$\scriptstyle L$}
            \put (75,6.5) {$\scriptstyle f_y$} 
            \put (75,29) {$\scriptstyle f_y$} 
            \put (6,18) {\sffamily\setlength{\fboxsep}{0pt}\colorbox{white}{\strut\bfseries\textcolor{black}{\begin{turn}{90}{$P(f_y/\bar{f}_y)$}\end{turn}}}} 
            \put (23,17) {\sffamily\setlength{\fboxsep}{0pt}\colorbox{white}{\strut\bfseries\textcolor{black}{\begin{turn}{90}{$\scriptstyle\bar{f}_y$}\end{turn}}}} 
            \put (86,32) {\sffamily\setlength{\fboxsep}{0pt}\colorbox{white}{\strut\bfseries\textcolor{black}{\begin{turn}{90}{$\scriptstyle P(f_y)$}\end{turn}}}} 
            \put (86,10) {\sffamily\setlength{\fboxsep}{0pt}\colorbox{white}{\strut\bfseries\textcolor{black}{\begin{turn}{90}{$\scriptstyle P(f_y)$}\end{turn}}}} 
            \put (54,41.5) {\sffamily\setlength{\fboxsep}{0pt}\colorbox{black}{\strut\bfseries\textcolor{white}{\small$(a)$}}} 
            \put (80,41.5) {\sffamily\setlength{\fboxsep}{0pt}\colorbox{black}{\strut\bfseries\textcolor{white}{\small$(b)$}}} 
            \put (80,19.0) {\sffamily\setlength{\fboxsep}{0pt}\colorbox{black}{\strut\bfseries\textcolor{white}{\small$(c)$}}} 
           \begin{tikzpicture}
                \coordinate (a) at (0,0); 
                \node[] at (a) {\tiny.};                 
                \coordinate (center1) at (7.8,2.9); 
				 \draw[black] (center1) circle (2pt);
                \coordinate (b) at ($ (center1) - (.075,0) $);
                \coordinate (c) at ($ (b) - (.1,0) $); 
                \coordinate (d) at ($ (center1) + (.075,0) $);
                \coordinate (e) at ($ (d) + (.1,0) $); 
                \draw[black][line width=0.1mm] (b) -- (c); 
                \draw[black][line width=0.1mm] (d) -- (e); 
			     \coordinate (center1) at (7.8,2.61); 
                \coordinate (b) at ($ (center1) - (.075,0) $);
                \coordinate (c) at ($ (b) - (.1,0) $); 
                \coordinate (d) at ($ (center1) + (.075,0) $);
                \coordinate (e) at ($ (d) + (.1,0) $); 
				 \draw[red] ($(b)-(0.0,0.07)$) rectangle ($ (d) + (0.0,0.07) $); 
                \draw[red][line width=0.1mm] (b) -- (c); 
                \draw[red][line width=0.1mm] (d) -- (e); 
			     \coordinate (center1) at (7.8,2.31); 
                \coordinate (b) at ($ (center1) + (0.075,0.0) $); 
                \coordinate (c) at ($ (center1) + (0,0.075) $); 
                \coordinate (d) at ($ (center1) + (-0.075,0) $);
                \coordinate (e) at ($ (center1) + (0.0,-0.075) $);
				 \draw[green] (b) -- (c) -- (d) -- (e) -- (b);
                \coordinate (c) at ($ (b) + (.1,0) $); 
                \coordinate (e) at ($ (d) - (.1,0) $); 
                \draw[green][line width=0.1mm] (b) -- (c); 
                \draw[green][line width=0.1mm] (d) -- (e); 
			     \coordinate (center1) at (7.8,2.01); 
                \coordinate (b) at ($ (center1) - 0.17*(0.5,0.289) $); 
                \coordinate (c) at ($ (center1) + 0.17*(0.5,-0.289) $); 
                \coordinate (d) at ($ (center1) + 0.17*(0,0.366) $);
				 \draw[blue] (b) -- (c) -- (d) -- (b);
                \coordinate (b) at ($ (center1) - (.075,0) $);
                \coordinate (c) at ($ (b) - (.1,0) $); 
                \coordinate (d) at ($ (center1) + (.075,0) $);
                \coordinate (e) at ($ (d) + (.1,0) $); 
                \draw[blue][line width=0.1mm] (b) -- (c); 
                \draw[blue][line width=0.1mm] (d) -- (e); 
				\coordinate (center1) at (3.22,1.75); 
				\coordinate (b) at ($ (center1) - 0.51*(1,0) $);
				\coordinate (c) at ($ (center1) - 0.25*(0,1) $); 
				\draw[red,line width=0.1mm] (center1) -- (b); 
				\draw[red,line width=0.1mm] (center1) -- (c); 
				\draw[red,line width=0.1mm] (b) -- (c); 
				\node at ($(center1)+(0.83,-0.15)$) {{\color{red}$\scriptstyle d-d_f(2-\tau)$}}; 
				\node at ($(center1)+(-0.2,0.15)$) {{\color{red}$\scriptstyle 1$}}; 
			            \end{tikzpicture}
                    \end{overpic}
                            \end{center}
    \caption{Statistical behavior of the failure threshold $f_y$ at different sizes $L=10, 20, 40, 80$. (\textbf{a}) Rescaled distributions $P(f_y/\bar{f}_y)$ plotted against $f_y/\bar{f}_y$ (\textbf{b}) Threshold distributions $P(f_y)$ on log-log scale. (\textbf{c}) $P(f_y)$ vs. $f_y$ on log-lin scale. The inset plots the mean threshold value $\bar{f}_y$ against $L$. The dashdotted (red) line is a guide to power-law $\bar{f}_y\propto 1/L^{d-d_f(2-\tau)}$ with $d=2$, $d_f=1.2$, and $\tau=1.3$.}
    \label{fig:pdfX}
\end{figure}
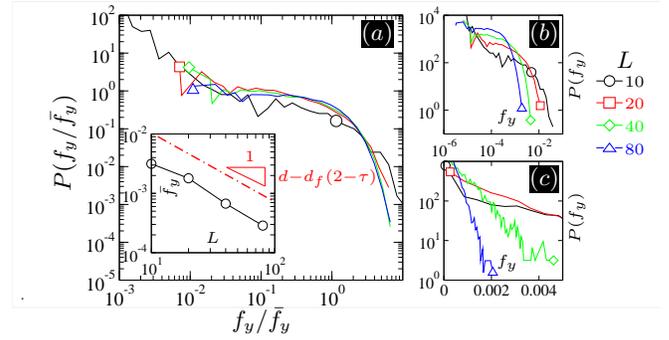
\section{``Wait time" distributions}\label{sec:WaitTimeDistributions}
Statistically speaking, the temporal characteristics of single particle diffusion should pertain to the intermittency of stress avalanches as is qualitatively seen in Fig.~\ref{fig:loadCurve}.
Stress drops tend to be well-coincided with intensely rearranging tracer particles.associat 
The \emph{quiescent} intervals in $u$, on the other hand, correspond closely with the accumulation periods of bulk stress. 

In the framework of the yielding transition, this interevent dynamics is commonly quantified via the instability threshold $f_y$.
Figure~\ref{fig:pdfX} quantifies fluctuations in this quantity at multiple system sizes.
In Fig.~\ref{fig:pdfX}(b) and (c), the exponential decay of our data, \emph{i.e.} $P(f_y)= \bar{f}_y^{-1}\text{exp}(-f_y/\bar{f}_y)$, suggests a Poissonian nature of the underlying yielding mechanism.
The data collapse of Fig.~\ref{fig:pdfX}(a) validates this hypothesis.
The inset of Fig.~\ref{fig:pdfX}(a) confirms that the mean stress threshold $\bar{f}_y$ decays with system size $L$. 
The associated scaling exponent follows from the stress conservation argument, \emph{i.e.} $\langle\dot\sigma\rangle=0$ which results in $\bar{f}_y=L^{-d}\langle S\rangle$.
The mean avalanche size is given by \cite{LinPNAS2014}
\begin{eqnarray} \label{eq:meanS}
\langle S\rangle&=&\int SP(S)~dS \nonumber \\ 
&\propto& L^{d_f(2-\tau)}~~~~~~1<\tau<2 ,
\end{eqnarray} 
and, therefore, $\langle f_y\rangle\propto 1/L^{d-d_f(2-\tau)}$.
Lower $\tau$ exponents (or equivalently higher frequency of larger events) tend to increase $\bar f_y$, in accordance with the proposed scaling, as the system will have to accommodate further stress to compensate larger avalanche-induced energy losses.  

The size dependence of the average instability distance could be understood in terms of the weakest link hypothesis. 
As required by the marginal stability criterion \cite{lin2014density}, the \emph{local} threshold spectrum associated with each microscopic constituent will vanish at extremely low threshold values, \emph{i.e.} $P(f_{\text{local}})\propto f_{\text{local}}^\theta$ with $\theta>0$.
Given the assumption of \emph{independence} between yielding events and minimal threshold criterion, the global stability condition is expected to follow Weibull statistics \cite{karmakar2010statistical,lin2014density}
\begin{equation}
P(f_y)=\frac{1}{ \bar{f}_y}(\frac{f_y}{\bar{f}_y})^\theta\text{exp}[-(\frac{f_y}{\bar{f}_y})^{\theta+1}],
\end{equation}
where $\bar{f}_y\propto 1/L^{\frac{d}{1+\theta}}$.

Within the plastic flow regime with $\theta\simeq 0$, Weibull statistics reduce to a Poisson point process which includes purely exponential interevent distributions as suggested by the numerics. 
However, the finite-size scaling associated with the yielding rate $\bar{f}_y^{-1}$ does not seem to simply follow from the above theoretical derivations $\bar{f}_y\propto 1/L^d$ which largely ignores the role of spatial correlations.   
Instead, the observed scaling can be generically rationalized from energy conservation principles.

\section{diffusive dynamics} \label{sec:diffusiveDynamics}
In previous sections, the emerging fluctuations, as a consequence of intermittent stress relaxations, were shown to include scaling characteristics that could be mainly interpreted in the plastic yielding framework. 
The long-term dynamics, however, may still enter a limiting Fickian regime which largely neglects short-lived correlation features leading to an effective Brownian process at infinite times. 

In order to validate the diffusion-based picture, we analyzed the total non-affine displacements $X=\sum_{i=1}^{N}u_i$ accumulated over $N$ loading steps.
Figure~\ref{fig:pdfUintegrated}(a) displays the temporal evolution of $X$ associated with several tracer particles. 
Such signals can be typically obtained by integrating over the fluctuating noise as in Fig.~\ref{fig:loadCurve}(b).
The inset of Fig.~\ref{fig:pdfUintegrated}(a) presents the power spectrum of the signal $\langle|\hat{X}(\omega)|^2\rangle$ decaying as $\omega^{-2}$, an expected scaling for a Brownian-type noise. 
In the inset of Fig.~\ref{fig:pdfUintegrated}(b), the probability distributions $P(X)$ are plotted at large strain values $\epsilon$. 
The rescaled data in Fig.~\ref{fig:pdfUintegrated}(b), $\epsilon^{\frac{1}{2}}P(X)$ vs. $X/\epsilon^{\frac{1}{2}}$, collapse on a master curve indicating a Brownian diffusion process at long times.

Fluctuations in $X$ will certainly depend on how frequent stress avalanches occur over the course of plastic flow. 
We further ignore tracers motion during elastic loading phase which are presumably small, compared to plastic deformations, making negligible contributions to the long-time particle diffusion.
Accordingly, $N$ may represent the total number of incurred events over a given strain interval $\epsilon$ and fluctuate in accordance with the Poisson distribution
\begin{equation}
P(N;\epsilon) = \frac{(\lambda\epsilon)^N}{N!}\text{exp}(-\lambda\epsilon),
\end{equation}
with the event rate $\lambda$ and mean number $\langle  N\rangle=\lambda\epsilon$.
Considering independent events, it follows that $\langle X^2|N\rangle=N\langle u^2\rangle$, and therefore $\langle X^2\rangle=\lambda\langle u^2\rangle\epsilon$.
We additionally assume that $\bar{f}_y$ controls the yielding rate, \emph{i.e.} $\lambda^{-1}\sim\bar{f}_y$. 
Inserting $\bar{f}_y\propto 1/L^{d-d_f(2-\tau)}$ and $\langle u^2\rangle\propto 1/L^{d_f(\tau-1)}$ gives $\langle X^2\rangle\propto L^{d-d_f}\epsilon$ with the \emph{effective} diffusion constant scaling as $D\propto L^{d-d_f}$, in line with the derivation in \cite{tyukodi2018diffusion}. 

Interestingly, $D$ will have no dependence on $\tau$, according to the derived relation, as this critical exponent makes equal contributions to the fluctuation size $\langle u^2\rangle$ and average yielding time $\bar f_y$.  
This will result in a diffusion coefficient that is only sensitive to the topology of triggered avalanches (\emph{i.e.} $d_f$ at a fixed system size $L$).
In other words, the size dependence will drop with uncorrelated avalanches filling up the entire space uniformly or equivalently $d\simeq d_f$. 

Figure \ref{fig:msd}(a) examines the linear growth of the mean squared displacements with $\epsilon$ within the steady-state flow regime.
At all system sizes, $\langle X^2\rangle$ exhibits a robust cross-over to the Fickian regime at large strains, \emph{i.e.} $\epsilon>0.1$.
The transition is more evident in Fig.~\ref{fig:msd}(b) with $\langle X^2\rangle/\epsilon$ reaching a size-dependent plateau as the deformation proceeds. 
The initial \emph{supper}-diffusive regime can be attributed to the correlated elastic-type deformation that a marginally stable system accommodates up to a size-dependent threshold $f_y$.
Figure~\ref{fig:msd}(c) shows size effects associated with the effective diffusion coefficient along with the proposed finite-size scaling that seems to be robust over at least one order of magnitude. 
\begin{figure}[t]
    \begin{center}
        \begin{overpic}[width=8.6cm]{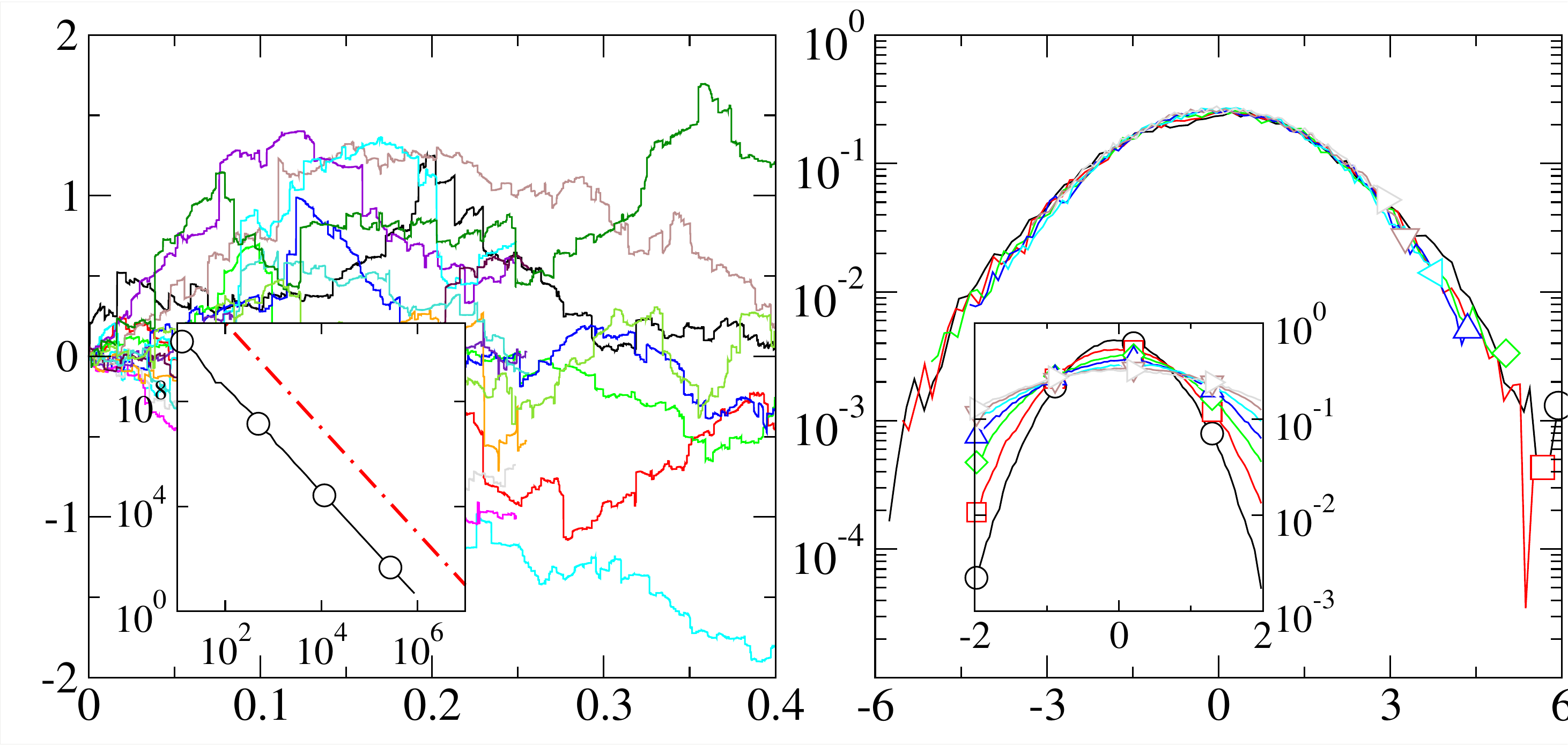}
			 \put (87,30) {$\epsilon$} 
            \put (88.8,26.6) {{\color{black}$\scriptstyle~0.09$}} 
            \put (88.8,23.1) {{\color{red}$\scriptstyle~0.12$}} 
            \put (88.8,19.8) {{\color{green}$\scriptstyle~0.16$}} 
            \put (88.8,16.2) {{\color{blue}$\scriptstyle~0.21$}} 
            \put (88.8,12.5) {{\color{cyan}$\scriptstyle~0.28$}} 
            \put (88.8,9.1) {{\color{brown}$\scriptstyle~0.37$}} 
            \put (88.8,5.8) {{\color{gray}$\scriptstyle~0.46$}} 
            \put (26,-3) {$\epsilon$}
            \put (74,-3) {$X/\epsilon^{\frac{1}{2}}$} 
            \put (70,10) {$\scriptstyle X$} 
            \put (20,10) {$\scriptstyle\omega$}
            \put (-4,22) {\sffamily\setlength{\fboxsep}{0pt}\colorbox{white}{\strut\bfseries\textcolor{black}{\begin{turn}{90}{$X$}\end{turn}}}} 
            \put (101,18) {\sffamily\setlength{\fboxsep}{0pt}\colorbox{white}{\strut\bfseries\textcolor{black}{\begin{turn}{90}{$\epsilon^{\frac{1}{2}}P(X)$}\end{turn}}}} 
            \put (59,14) {\sffamily\setlength{\fboxsep}{0pt}\colorbox{white}{\strut\bfseries\textcolor{black}{\begin{turn}{90}{$\scriptstyle P(X)$}\end{turn}}}} 
            \put (12.5,9.9) {\sffamily\setlength{\fboxsep}{0pt}{\strut\bfseries\textcolor{black}{\begin{turn}{90}{$\scriptstyle \langle|\hat{X}(\omega)|^2\rangle$}\end{turn}}}} 
            \put (44,41.5) {\sffamily\setlength{\fboxsep}{0pt}\colorbox{black}{\strut\bfseries\textcolor{white}{\small$(a)$}}} 
            \put (94,41.5) {\sffamily\setlength{\fboxsep}{0pt}\colorbox{black}{\strut\bfseries\textcolor{white}{\small$(b)$}}} 
           \begin{tikzpicture}
                \coordinate (a) at (0,0); 
                \node[] at (a) {\tiny.};                 
                \coordinate (center1) at (7.4,2.23); 
				 \draw[black] (center1) circle (2pt);
                \coordinate (b) at ($ (center1) - (.075,0) $);
                \coordinate (c) at ($ (b) - (.1,0) $); 
                \coordinate (d) at ($ (center1) + (.075,0) $);
                \coordinate (e) at ($ (d) + (.1,0) $); 
                \draw[black][line width=0.1mm] (b) -- (c); 
                \draw[black][line width=0.1mm] (d) -- (e); 
                \coordinate (center1) at (7.4,1.93); 
                \coordinate (b) at ($ (center1) - (.075,0) $);
                \coordinate (c) at ($ (b) - (.1,0) $); 
                \coordinate (d) at ($ (center1) + (.075,0) $);
                \coordinate (e) at ($ (d) + (.1,0) $); 
				 \draw[red] ($(b)-(0.0,0.07)$) rectangle ($ (d) + (0.0,0.07) $); 
                \draw[red][line width=0.1mm] (b) -- (c); 
                \draw[red][line width=0.1mm] (d) -- (e); 
                \coordinate (center1) at (7.4,1.63); 
                \coordinate (b) at ($ (center1) + (0.075,0.0) $); 
                \coordinate (c) at ($ (center1) + (0,0.075) $); 
                \coordinate (d) at ($ (center1) + (-0.075,0) $);
                \coordinate (e) at ($ (center1) + (0.0,-0.075) $);
				 \draw[green] (b) -- (c) -- (d) -- (e) -- (b);
                \coordinate (c) at ($ (b) + (.1,0) $); 
                \coordinate (e) at ($ (d) - (.1,0) $); 
                \draw[green][line width=0.1mm] (b) -- (c); 
                \draw[green][line width=0.1mm] (d) -- (e); 
                \coordinate (center1) at (7.4,1.33); 
                \coordinate (b) at ($ (center1) - 0.17*(0.5,0.289) $); 
                \coordinate (c) at ($ (center1) + 0.17*(0.5,-0.289) $); 
                \coordinate (d) at ($ (center1) + 0.17*(0,0.366) $);
				 \draw[blue] (b) -- (c) -- (d) -- (b);
                \coordinate (b) at ($ (center1) - (.075,0) $);
                \coordinate (c) at ($ (b) - (.1,0) $); 
                \coordinate (d) at ($ (center1) + (.075,0) $);
                \coordinate (e) at ($ (d) + (.1,0) $); 
                \draw[blue][line width=0.1mm] (b) -- (c); 
                \draw[blue][line width=0.1mm] (d) -- (e); 
			     \coordinate (center1) at (7.4,1.03); 
                \coordinate (b) at ($ (center1) + 0.17*(0.289,0.5) $); 
                \coordinate (c) at ($ (center1) + 0.17*(0.289,-0.5) $); 
                \coordinate (d) at ($ (center1) + 0.17*(-0.366,0.0) $);
				 \draw[cyan] (b) -- (c) -- (d) -- (b);
                \coordinate (b) at ($ (center1) - (.075,0) $);
                \coordinate (c) at ($ (b) - (.1,0) $); 
                \coordinate (d) at ($ (center1) + (.075,0) $);
                \coordinate (e) at ($ (d) + (.1,0) $); 
               	 \draw[cyan][line width=0.1mm] (b) -- (c); 
                	\draw[cyan][line width=0.1mm] (d) -- (e); 
			     \coordinate (center1) at (7.4,0.73); 
                \coordinate (b) at ($ (center1) - 0.17*(0.5,-0.289) $); 
                \coordinate (c) at ($ (center1) + 0.17*(0.5,+0.289) $); 
                \coordinate (d) at ($ (center1) - 0.17*(0,0.366) $);
				 \draw[brown] (b) -- (c) -- (d) -- (b);
                \coordinate (b) at ($ (center1) - (.075,0) $);
                \coordinate (c) at ($ (b) - (.1,0) $); 
                \coordinate (d) at ($ (center1) + (.075,0) $);
                \coordinate (e) at ($ (d) + (.1,0) $); 
                \draw[brown][line width=0.1mm] (b) -- (c); 
                \draw[brown][line width=0.1mm] (d) -- (e); 
			     \coordinate (center1) at (7.4,0.43); 
                \coordinate (b) at ($ (center1) + 0.17*(-0.289,0.5) $); 
                \coordinate (c) at ($ (center1) + 0.17*(-0.289,-0.5) $); 
                \coordinate (d) at ($ (center1) + 0.17*(0.366,0.0) $);
				 \draw[gray] (b) -- (c) -- (d) -- (b);
                \coordinate (b) at ($ (center1) - (.075,0) $);
                \coordinate (c) at ($ (b) - (.1,0) $); 
                \coordinate (d) at ($ (center1) + (.075,0) $);
                \coordinate (e) at ($ (d) + (.1,0) $); 
                \draw[gray][line width=0.1mm] (b) -- (c); 
                \draw[gray][line width=0.1mm] (d) -- (e); 
				\coordinate (center1) at (2.1,1.6); 
				\coordinate (b) at ($ (center1) - 0.37*(1,0) $);
				\coordinate (c) at ($ (center1) - 0.4*(0,1) $); 
				\draw[red,line width=0.1mm] (center1) -- (b); 
				\draw[red,line width=0.1mm] (center1) -- (c); 
				\draw[red,line width=0.1mm] (b) -- (c); 
				\node at ($(center1)+(0.12,-0.15)$) {{\color{red}$\scriptstyle 2$}}; 
				\node at ($(center1)+(-0.2,0.13)$) {{\color{red}$\scriptstyle 1$}}; 
				\end{tikzpicture}
                    \end{overpic}
                            \end{center}
    \caption{Dynamics of the integrated noise $X=\sum_{i=1}^{N}u_i$ and relevant statistics at $L=80$. (\textbf{a}) Accumulated displacements $X$ incurred over the imposed strain $\epsilon$ corresponding to multiple tracer particles. The inset illustrates the associated power spectrum $\langle|\hat{X}(\omega)|^2\rangle$ in the frequency domain $\omega$. The dashdotted line indicates $\langle|\hat{X}(\omega)|^2\rangle\propto\omega^{-2}$.  (\textbf{b}) Rescaled probability distributions $\epsilon^{\frac{1}{2}}P(X)$ as a function of $X/\epsilon^{\frac{1}{2}}$ at different $\epsilon$. The inset shows the unrescaled data.}
    \label{fig:pdfUintegrated}
\end{figure}
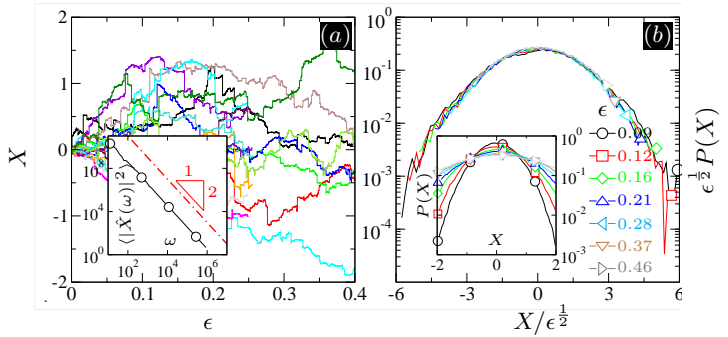
\begin{figure}[t]
    \begin{center}
        \begin{overpic}[width=8.6cm]{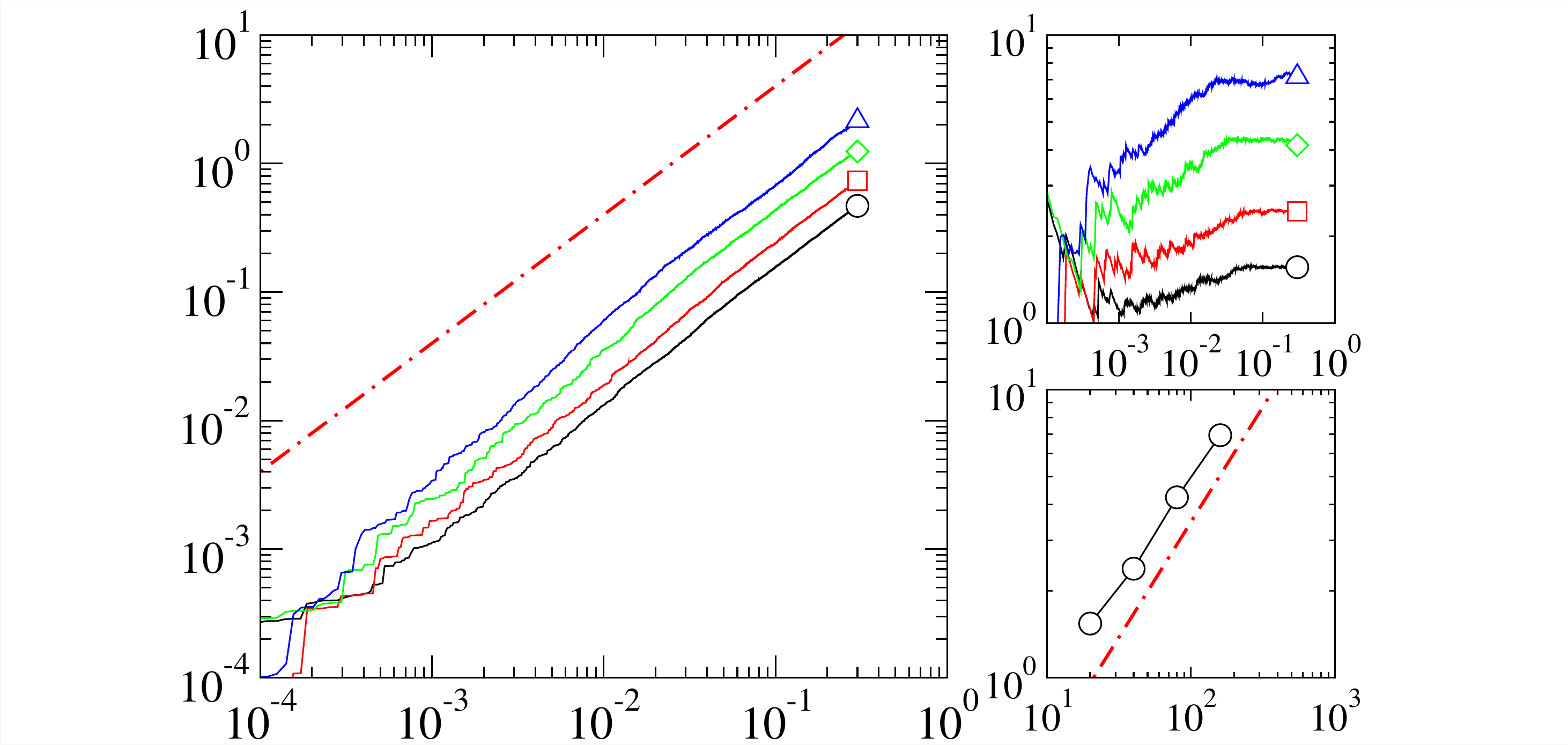}
			 \put (53.5,21.2) {$L$} 
            \put (53.5,18.25) {{\color{black}$\scriptstyle~20$}} 
            \put (53.5,14.8) {{\color{red}$\scriptstyle~40$}} 
            \put (53.5,11.3) {{\color{green}$\scriptstyle~80$}} 
            \put (53.5,7.8) {{\color{blue}$\scriptstyle~160$}} 
            \put (38,-3) {$\epsilon$}
            \put (75,-2.7) {$\scriptstyle L$} 
            \put (76,29) {\sffamily\setlength{\fboxsep}{0pt}\colorbox{white}{\strut\bfseries\textcolor{black}{{$\epsilon$}}}} 
            \put (6,21) {\sffamily\setlength{\fboxsep}{0pt}\colorbox{white}{\strut\bfseries\textcolor{black}{\begin{turn}{90}{$\langle X^2\rangle$}\end{turn}}}} 
            \put (86,32) {\sffamily\setlength{\fboxsep}{0pt}\colorbox{white}{\strut\bfseries\textcolor{black}{\begin{turn}{90}{$\scriptstyle\langle X^2\rangle/\epsilon$}\end{turn}}}} 
            \put (87,13) {\sffamily\setlength{\fboxsep}{0pt}\colorbox{white}{\strut\bfseries\textcolor{black}{\begin{turn}{90}{$\scriptstyle D$}\end{turn}}}} 
            \put (18,41.5) {\sffamily\setlength{\fboxsep}{0pt}\colorbox{black}{\strut\bfseries\textcolor{white}{\small$(a)$}}} 
            \put (68,41.5) {\sffamily\setlength{\fboxsep}{0pt}\colorbox{black}{\strut\bfseries\textcolor{white}{\small$(b)$}}} 
            \put (68,19.0) {\sffamily\setlength{\fboxsep}{0pt}\colorbox{black}{\strut\bfseries\textcolor{white}{\small$(c)$}}} 
           \begin{tikzpicture}
                \coordinate (a) at (0,0); 
                \node[] at (a) {\tiny.};                 
                \coordinate (center1) at (4.3,1.5); 
				 \draw[black] (center1) circle (2pt);
                \coordinate (b) at ($ (center1) - (.075,0) $);
                \coordinate (c) at ($ (b) - (.1,0) $); 
                \coordinate (d) at ($ (center1) + (.075,0) $);
                \coordinate (e) at ($ (d) + (.1,0) $); 
                \draw[black][line width=0.1mm] (b) -- (c); 
                \draw[black][line width=0.1mm] (d) -- (e); 
			     \coordinate (center1) at (4.3,1.21); 
                \coordinate (b) at ($ (center1) - (.075,0) $);
                \coordinate (c) at ($ (b) - (.1,0) $); 
                \coordinate (d) at ($ (center1) + (.075,0) $);
                \coordinate (e) at ($ (d) + (.1,0) $); 
				 \draw[red] ($(b)-(0.0,0.07)$) rectangle ($ (d) + (0.0,0.07) $); 
                \draw[red][line width=0.1mm] (b) -- (c); 
                \draw[red][line width=0.1mm] (d) -- (e); 
			     \coordinate (center1) at (4.3,0.91); 
                \coordinate (b) at ($ (center1) + (0.075,0.0) $); 
                \coordinate (c) at ($ (center1) + (0,0.075) $); 
                \coordinate (d) at ($ (center1) + (-0.075,0) $);
                \coordinate (e) at ($ (center1) + (0.0,-0.075) $);
				 \draw[green] (b) -- (c) -- (d) -- (e) -- (b);
                \coordinate (c) at ($ (b) + (.1,0) $); 
                \coordinate (e) at ($ (d) - (.1,0) $); 
                \draw[green][line width=0.1mm] (b) -- (c); 
                \draw[green][line width=0.1mm] (d) -- (e); 
			     \coordinate (center1) at (4.3,0.61); 
                \coordinate (b) at ($ (center1) - 0.17*(0.5,0.289) $); 
                \coordinate (c) at ($ (center1) + 0.17*(0.5,-0.289) $); 
                \coordinate (d) at ($ (center1) + 0.17*(0,0.366) $);
				 \draw[blue] (b) -- (c) -- (d) -- (b);
                \coordinate (b) at ($ (center1) - (.075,0) $);
                \coordinate (c) at ($ (b) - (.1,0) $); 
                \coordinate (d) at ($ (center1) + (.075,0) $);
                \coordinate (e) at ($ (d) + (.1,0) $); 
                \draw[blue][line width=0.1mm] (b) -- (c); 
                \draw[blue][line width=0.1mm] (d) -- (e); 
				\coordinate (center1) at (2.4,2.7); 
				\coordinate (b) at ($ (center1) + 0.52*(1,0) $);
				\coordinate (c) at ($ (center1) - 0.36*(0,1) $); 
				\draw[red,line width=0.1mm] (center1) -- (b); 
				\draw[red,line width=0.1mm] (center1) -- (c); 
				\draw[red,line width=0.1mm] (b) -- (c); 
				\node at ($(center1)+(0.2,0.13)$) {{\color{red}$\scriptstyle 1$}}; 
				\node at ($(center1)+(-0.1,-0.1)$) {{\color{red}$\scriptstyle 1$}}; 
				\coordinate (center1) at (6.5,0.6); 
				\coordinate (b) at ($ (center1) - 0.3*(1,0) $);
				\coordinate (c) at ($ (center1) + 0.45*(0,1) $); 
				\draw[red,line width=0.1mm] (center1) -- (b); 
				\draw[red,line width=0.1mm] (center1) -- (c); 
				\draw[red,line width=0.1mm] (b) -- (c); 
				\node at ($(center1)+(0.36,0.2)$) {{\color{red}$\scriptstyle d-d_f$}}; 
				\node at ($(center1)+(-0.15,-0.15)$) {{\color{red}$\scriptstyle 1$}}; 
				\end{tikzpicture}
                    \end{overpic}
                            \end{center}
    \caption{Temporal evolution of the mean squared displacement at multiple size $L$ (\textbf{a}) $\langle X^2\rangle$ against $\epsilon$. The dashdotted line with slope $1$ indicates the linear growth $\langle X^2\rangle\propto \epsilon$. (\textbf{b}) Rescaled quantity $\langle X^2\rangle/\epsilon$ versus $\epsilon$. (\textbf{c}) Size-dependence of the effective diffusion coefficient $D$. The dashdotted line is a guide to power law $D\propto L^{d-d_f}$ with $d=2$ and $d_f=1.2$.}
    \label{fig:msd}
\end{figure}
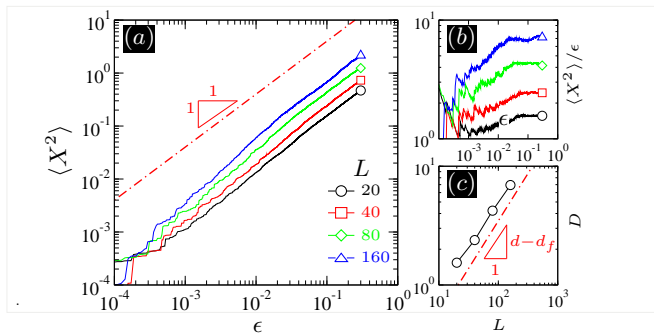

\section{Conclusions} 
We have analyzed the trajectories of individual particles in driven soft amorphous solids that exhibit scale-dependent fluctuation features both in terms of fluctuation amplitudes and occurrence frequencies.
The observed properties closely resemble those of stress fluctuations that are commonly attributed to spatially extended stress avalanches inducing long-range deformation features within the medium.
In this framework, ``jump" size fluctuations are \emph{kinematically} quantified based on the concept of Eshelby shear transformations.

The temporal dynamics, on the other hand, was described by introducing the stress instability threshold $f_y$ that is, on average, related to the mean avalanche size, as a result of the energy conservation principle.
Fluctuations in $f_y$ reveal a fast exponential-like decay which is at odds with broad scale-free avalanche-size distributions.
The former exhibits a size-dependent characteristic stress scale $\bar{f}_y$ that must be linked with local energy ``cages" in glassy structures \cite{lechenault2010super}.
The meta-stability hypothesis implies that fluctuations in local energy thresholds will be bounded as the driven solid always remain in \emph{near}-critical states within the plastic flow regime \cite{lin2014density}.
Recent analysis of local yield stresses in sheared glasses made by Patinet \emph{et al.} \cite{patinet2016connecting} led to narrow threshold distributions which is in agreement with this stability argument.

The observed self-diffusion can be understood in terms of ``fading" temporal correlations in the long-time limit that emerge as a consequence of disorder and heterogeneities which prevails the short-term collective dynamics in space.
This argument appears to be in line with the Poisson-type activation process within the plastic flow regime.
Spatial correlations, however, control displacement fluctuations as well as the yielding rate leading to finite-size scaling in the diffusion factor.

\emph{Acknowledgment-} We acknowledge insightful discussions with J-L. Barrat, K. Martens, G. Gradenigo, and E. Bertin about our results.     
\newpage
\bibliography{ref}
\end{document}